\documentclass[%
 reprint,
 amsmath,amssymb,
 aps,
]{revtex4-2}

\usepackage{orcidlink}
\usepackage{graphicx}
\usepackage{dcolumn}
\usepackage{bm}

\begin{document}

\preprint{APS/123-QED}

\title{Widefield Quantum Sensor for Vector Magnetic Field Imaging of Micromagnetic Structures}

\author{Orlando D. Cunha \orcidlink{0009-0008-1608-9034}}
\affiliation{INL - International Iberian Nanotechnology Laboratory, Nieder Group on Ultrafast Bio- and Nanophotonics, 4715-330 Braga, Portugal}
\author{Filipe Camarneiro \orcidlink{0000-0003-3102-2205}}
\affiliation{INL - International Iberian Nanotechnology Laboratory, Nieder Group on Ultrafast Bio- and Nanophotonics, 4715-330 Braga, Portugal}
\author{João P. Silva \orcidlink{0000-0003-1180-2941}}
\affiliation{INL - International Iberian Nanotechnology Laboratory, Nieder Group on Ultrafast Bio- and Nanophotonics, 4715-330 Braga, Portugal}
\author{Hariharan Nhalil \orcidlink{0000-0003-4970-7106}}
\affiliation{Department of Physics, Institute of Nanotechnology and Advanced Materials, Bar-Ilan University, 5290002 Ramat-Gan, Israel}
\author{Ariel Zaig \orcidlink{0000-0001-6639-5234}}
\affiliation{Department of Physics, Institute of Nanotechnology and Advanced Materials, Bar-Ilan University, 5290002 Ramat-Gan, Israel}
\author{Lior Klein \orcidlink{0000-0002-1233-9453}}
\affiliation{Department of Physics, Institute of Nanotechnology and Advanced Materials, Bar-Ilan University, 5290002 Ramat-Gan, Israel}
\author{Jana B. Nieder \orcidlink{0000-0002-4973-1889}}
\email{Corresponding author: jana.nieder@inl.int}
\affiliation{INL - International Iberian Nanotechnology Laboratory, Nieder Group on Ultrafast Bio- and Nanophotonics, 4715-330 Braga, Portugal}

\date{\today}

\begin{abstract}
Many spintronic, magnetic-memory, and neuromorphic devices rely on spatially varying magnetic fields. Quantitatively imaging these fields with full vector information over extended areas remains a major challenge. Existing probes either offer nanoscale resolution at the cost of slow scanning, or widefield imaging with limited vector sensitivity or material constraints. Quantum sensing with nitrogen-vacancy (NV) centers in diamond promises to bridge this gap, but a practical camera-based vector magnetometry implementation on relevant microstructures has not been demonstrated. Here we adapt a commercial widefield microscope to implement a camera-compatible pulsed optically detected magnetic resonance protocol to reconstruct stray-field vectors from microscale devices. By resolving the Zeeman shifts of the four NV orientations, we reconstruct the stray-field vector generated by microfabricated permalloy structures that host multiple stable remanent states. Our implementation achieves a spatial resolution of $\approx 0.52 ~\mu\mathrm{m}$ across an $83~\mu\mathrm{m} \times 83~\mu\mathrm{m}$ field of view and a peak sensitivity of $ (828 \pm 142)~\mathrm{nT\,Hz^{-1}}$, with acquisition times of only a few minutes. These results establish pulsed widefield NV magnetometry on standard microscopes as a practical and scalable tool for routine vector-resolved imaging of complex magnetic devices.
\end{abstract}

\maketitle


\section{\label{sec:intro}Introduction}

Micromagnetic textures serve as the foundation for a wide range of emerging spintronic technologies, including memory devices \cite{Bhatti}, skyrmion-based logic \cite{ZhouY} and neuromorphic architectures \cite{JZhou} that exploit multistable magnetic states. In all of these systems, the relevant information is encoded not only in the presence of magnetic fields but also in their spatial distribution. Quantitatively mapping these fields with micrometer-scale spatial resolution, over large areas and with full vector information is therefore essential for understanding and optimizing these devices.

Several established techniques address parts of this problem, but each one comes with its own trade-offs. Electrical probes such as harmonic Hall measurements \cite{Kim,Hayashi} quantify effective spin–orbit torques and anisotropy fields, but they average over the device area and do not provide spatially resolved field maps. X-ray magnetic circular dichroism (XMCD) \cite{Bonetti,Donnelly} can image complex magnetization patterns, yet they require access to synchrotron radiation.  Magneto-optical Kerr effect (MOKE) microscopy \cite{McCord,Cao} offers fast, widefield imaging of domains and dynamics, but it is limited to materials with a strong Kerr response and usually yields only partial vector information. As a result, none of these methods provides a quantitative map of the full magnetic-field vector.

More recently, nitrogen-vacancy (NV) centers in diamond have emerged as one of the most promising quantum sensing platforms \cite{DOHERTY20131}. Their exceptional coherence properties at room temperature, optical addressability, and ability to reconstruct full magnetic vector fields make them ideal for investigating nano- and microscale magnetic phenomena \cite{Boretti,Xu}. In their negatively charged state, NV centers have an electronic spin triplet ground state ($S=1$). The Zeeman splitting of the $m_s=\pm1$ spin states encodes the projection of the magnetic field along a crystal axis, which can be readout using optically detected magnetic resonance (ODMR). Since NV centers can align along four distinct $\langle 111 \rangle$ directions within the crystal, measuring the Zeeman shifts associated with each orientation allows for the reconstruction of the full three-dimensional magnetic field vector at the NV imaging plane. 

\begin{figure}
\includegraphics{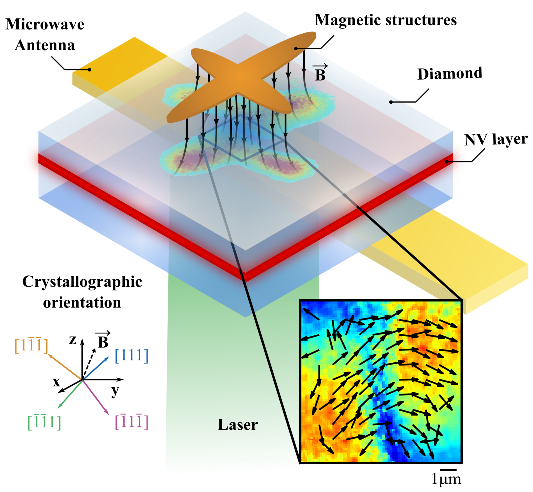}
\caption{\label{fig:im1} Illustration of nitrogen-vacancy (NV) vector magnetometry for imaging magnetic structures. A near-surface NV layer in diamond senses the stray field from microstructures, with each NV orientation measuring a different projection and enabling reconstruction of the magnetic-field vector.}
\end{figure}

This principle has been explored with single NV centers in both scanning-probe and confocal implementations \cite{Tetienne,WangG}. These approaches provide nanoscale spatial resolution, but they require point-by-point scans and often demand more complex setups or protocols, especially for full vector reconstruction \cite{Huxter}. In contrast, when an ensemble of near-surface NV centers is combined with widefield imaging, every camera pixel effectively becomes an independent vector magnetometer, enabling parallel readout of the magnetic field over thousands of spatial points in a single acquisition \cite{Nowodzinski,Turner}. While this unlocks the potential for rapid wide-field imaging, the technique has thus far been limited to continuous-wave ODMR protocols and applied largely to simple, proof-of-principle samples \cite{Chipaux2015,PhamL.M.,TetienneJ.P}.

In this work, we implement a camera-based pulsed ODMR protocol on a commercial widefield microscope, using a shallow implanted NV layer in bulk diamond to reconstruct the stray-field vectors generated by permalloy films (Figure \ref{fig:im1}). By fitting the ODMR spectra from the four NV orientations across the field of view, we extract the magnetic-field vector at the NV imaging plane. Beyond quantifying the spatial resolution and sensitivity of this platform, our results show that ensembles of NV centers provide a practical and accessible approach for the vector characterization of complex magnetic devices.

\section{Experimental methods}
\subsection{Nitrogen-vacancy center in diamond}
The NV center is a point defect in the diamond lattice consisting of a substitutional nitrogen neighboring a vacancy. Its negatively charged state forms a S=1 spin system with a ground state composed of the spin sublevel $m_\mathrm{s}=0$ and the degenerate $m_\mathrm{s}=\pm1$ (see energy-level scheme in Figure \ref{fig:2} a). Without the presence of a magnetic field and at room temperature, the $m_\mathrm{s}=\pm1$ are split from the $m_\mathrm{s}=0$ by D = 2.87 GHz. The spin-state-dependent intersystem crossing \cite{Goldman,Goldman2}, combined with its non-radiative and non-conserving spin transition, allows the initialization into the $m_\mathrm{s}=0$ state and the readout of the spin state through their optical emission using ODMR techniques \cite{Wrachtrup}. In the presence of an external magnetic field, the degeneracy of the $m_\mathrm{s}=\pm1$ sublevels is lifted due to the Zeeman effect, as illustrated in the inset of Figure \ref{fig:2} a). This split is proportional to the projection of the magnetic field strength and can be expressed as \cite{Rondin}:
\begin{equation}
B = \Delta v / 2\gamma
\end{equation}
where $\Delta v$ is the difference in frequency between the $m_\mathrm{s}=+1$ and the $m_\mathrm{s}=-1$, and $\gamma \approx$ 28 $\mathrm{GHz/T}$ is the NV gyromagnetic ratio. The vector reconstruction capability arises from the $C_{3\mathrm{v}}$ symmetry property of the NV centers, which allows their alignment along one of the four possible [111] directions of the diamond. Since each NV center senses only the magnetic field projection along its quantization axis, by analyzing the Zeeman shifts from the four crystallographic orientations, it is possible to reconstruct the full 3D magnetic vector field \cite{Chipaux2015}.
\subsection{Microstructured magnetic samples}
\begin{figure*}
\includegraphics{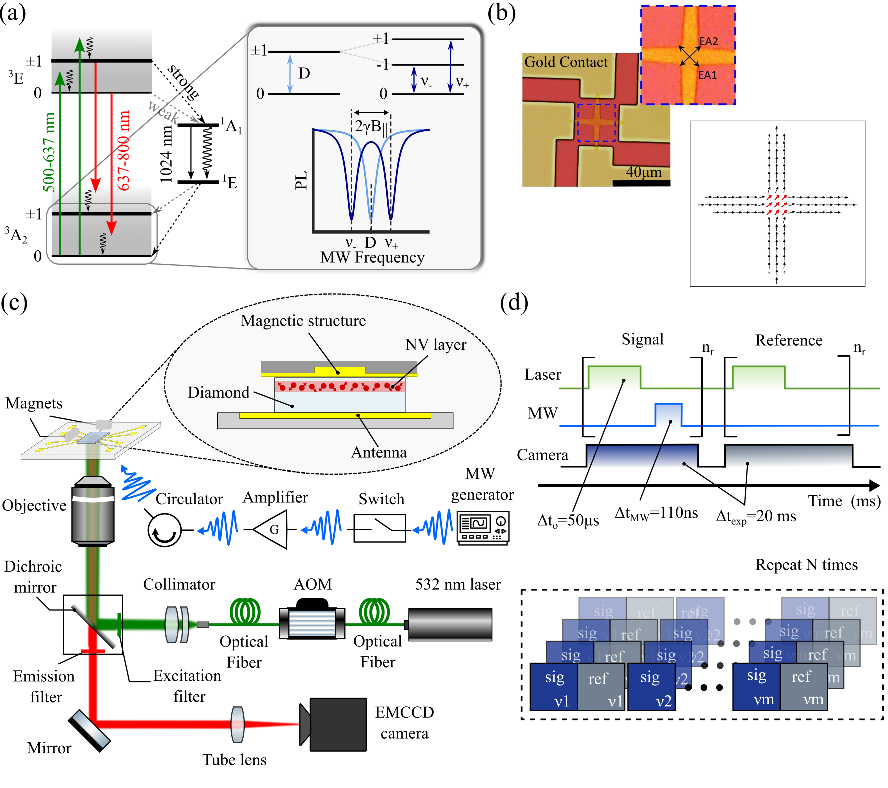}
\caption{\label{fig:2}Experimental setup for widefield NV-based magnetometry. (a) Energy level structure of the NV center with a triplet ground and excited state, and two intermediate states. Relevant transitions are highlighted by arrows. Inset: splitting of the magnetic spin states $m_\mathrm{s}=\pm1$ due to the Zeeman effect. (b) Optical microscope image of two crossing ellipse structures (left) and an Object Oriented MicroMagnetic Framework simulation of one of the four possible remanent states (right). (c) Schematic of the experimental NV-based setup. Inset: cross-section showing the sample configuration with magnetic structures placed above the diamond and microwave antenna. (d) Pulsed ODMR protocol for widefield imaging. Each sequence is repeated $\mathrm{n}_\mathrm{r}$ times to match the exposure time of the camera (top). Sequence of signal acquisition alternating between Signal and Reference protocols, with the microwave frequency sweep from point 1 to $m$, repeated $N$ times (bottom).}
\end{figure*}

The multilayer stack of Ta (5 nm)/NiFe (100 nm)/Ta (5 nm) was deposited on a naturally oxidized Si wafer employing an Ion Beam Sputtering (IBS) unit. The patterning of two-crossing ellipse structures was carried out using a combination of photolithography, ion milling, and wet etching. Subsequently, a gold layer approximately 1.5 times thicker than the NiFe layer was deposited, patterned, and lifted off to form the electrical contact leads and pads. An exemplary optical image of the fabricated structure is shown in Figure \ref{fig:2} (b). Each ellipse measures 40 $\mathrm{\mu m}$ in length and 5 $\mathrm{\mu m}$ in width, having an aspect ratio of 1:8. Due to magnetic shape anisotropy induced by the ellipses, the overlapping region exhibits an effective biaxial magnetic anisotropy, characterized by two orthogonal easy axes oriented at 45 ° relative to the major axes of the ellipses. The directions of the easy axes are marked in the inset panel of Figure \ref{fig:2} (b). This structure supports four stable magnetic remanent states \cite{Telepinsky2012}, and the right panel shows a micromagnetic simulation snapshot that illustrates the magnetic configuration of one of these states taken from \cite{Telepinsky2016}. The magnetic biaxial anisotropy field is calculated to be $\sim$ 30 mT, based on magnetic characterization measurements and fitting with the Stoner-Wohlfarth model for coherent magnetization rotation in the presence of biaxial magnetic anisotropy \cite{Telepinsky2016}. 

\subsection{Optical quantum vector magnetometry setup}
Figure \ref{fig:2} (c) schematically represents the commercial widefield/TIRF microscope integrated with additional hardware necessary to perform our ODMR experiments. In this work, a diamond slab with a shallow NV layer was commercially acquired from QNAMI. The diamond is a quantum-grade single crystal produced by chemical vapor deposition and cut along the [100] orientation. It has an area of 3 mm x 3 mm and a thickness of 50 $\mu m$, with an NV center density of around 1000 NVs/ $\mu m^2$. The NV centers are located about $\approx$ 10 nm below the top surface and are statistically distributed among the four [111] crystallographic directions. To excite the NV centers, we used a continuous-wave (CW) 532 nm laser (Roithner Lasertechnik GmbH 532-100-1 CD41932) coupled to an acousto-optic modulator (AOM) with an output power of 50 mW. The light was focused on the back aperture of the TIRF 100X oil objective (Nikon CFI Apochromat) over a field of view (FOV) of $\approx$ 82.75 $ \mu$m x 82.75 $ \mu$m before illuminating the sample. The emitted fluorescence from the NV layer is collected through the same objective and filtered using a G-2A filter combination (Nikon) before being digitally captured using a tube lens onto an electron multiplying charge coupled device camera (Andor iXon 897). Microwave signals are generated using the Rhode \& Schwarz (SMC100A) signal generator and connected to a MW switch (Mini-Circuits ZASWA-2-50DRA+), which enables control over the output of the MW signal via transistor-transistor logic (TTL). After the switch, the MW signal is amplified using a Mini-Circuits ZHL-16W-43-S+ amplifier, then passed through a circulator before being delivered to the sample via a linear MW antenna. To control the timing of the experiment's data acquisition and the gating of the laser and MW source, we used the SpinCore PulseBlaster (ESR-PRO 500 MHz). Two neodymium magnets were positioned around the diamond to create a bias field capable of splitting the degeneracy of the four NV orientations and ensuring that their contributions were independently measured. To identify the remanent state of the ellipses, the magnetic structures were positioned on top of the diamond and the cover glass (antenna) in a downward configuration, ensuring proximity to the NV sensing layer, as shown in the inset of Figure \ref{fig:2} (c). The acquired ODMR data were then transferred to a computer for post-processing and analysis using custom Python 3.10. scripts, enabling efficient data interpretation and visualization. 
\subsection{Vector magnetometry experimental procedure and readout sequence}
To compensate for the longer integration times of the camera, we employed an averaging scheme similar to Horsley \textit{et al.} \cite{Horsley} that repeats the pulse sequence multiple times ($\mathrm{n}_\mathrm{r}$) within a single camera exposure time (Figure \ref{fig:2} (d)). For each measurement point, a signal image was formed with the MWs on, followed by an image without the MWs to form a reference image. These two images were then combined pixel by pixel to calculate the Michelson contrast, which shows the change in fluorescence caused by the MW pulse. The contrast is defined as $\mathrm{C} = 1-\mathrm{C}_\mathrm{Signal}/ \mathrm{C}_\mathrm{Reference}$, where C corresponds to the fluorescence counts for each type of image. This approach mitigates random noise and fluctuations during measurements. Each measurement point was then averaged over a total of N repetitions to increase the signal-to-noise-ratio (SNR) and subsequently the sensitivity.   

Before running the pulsed ODMR sequences, we characterized the timing delays between the PulseBlaster output, the AOM and the microwave switch. The relative offsets, on the order of a few tens of nanoseconds, were corrected in the PulseBlaster sequence such that MW and laser pulses did not overlap. Following this protocol, we implemented the pulsed ODMR sequence as illustrated in the right side of Figure \ref{fig:2} (d). The sequence consists of $\Delta \mathrm{t}_\mathrm{o}= $ 50 $\mu$s laser pulse that serves as initialization and readout of the spin state, followed by a MW pulse of duration $\pi$ ($\Delta \mathrm{t}_\mathrm{MW}= $110 ns). This duration was obtained through a Rabi experiment for the \([1\bar{1}\bar{1}]\) orientation (2.7305 GHz). This sequence of laser and MWs is then repeated $\mathrm{n}_\mathrm{r}\approx$ 400 times to match the exposure time of the camera ($\Delta \mathrm{t}_\mathrm{exp}= $ 20 ms). The MW frequency was swept from 2.65 GHz to 2.88 GHz with a frequency step of 1 MHz and each measurement was repeated N=30 times, yielding a total acquisition time of 4.62 minutes. We chose to sweep only half of the ODMR spectrum since it allows faster experiments while retrieving the same information as the full ODMR spectrum, with the simple assumption that the spectrum is symmetric at 2.87 GHz.
\subsection{Sensitivity}
The sensitivity of these experiments is defined as the smallest change in magnetic field that can be reliably detected. A lower value corresponds to a better sensitivity, since it allows smaller magnetic field fluctuations to be detected. The optimized sensitivity $\eta$ for a shot-noise limited pulsed ODMR for an ensemble of NV centers is given by \cite{Barry}:
\begin{equation}\label{eq:n_p1}
    \eta = \frac{8}{3\sqrt{3}}\frac{\hbar}{g_e\mu_\mathrm{B}}\frac{1}{C\sqrt{\mathcal{N}}}\frac{\sqrt{t_\mathrm{o}+T_2^*}}{T_2^*}
\end{equation}
where $\hbar$ is the reduced Planck constant, $ g_e$ is the electron spin g-factor, $\mu_\mathrm{B}$ is the Bohr magneton, $C$ is the contrast, $\mathcal{N}$ is the mean number of photons collected per optical readout cycle, $t_\mathrm{o}$ is the overhead time, and $T_2^*$ is the dephasing time. The latter can be approximated by $T_2^*=1/(\pi\Gamma_\mathrm{p}) $ with $\Gamma_\mathrm{p}$ being the fitted linewidth of the ODMR dip.

In camera-based approaches, the quantity $\mathcal{N}$ is neither directly provided nor easily calculated. To address this, we adopted the formulation proposed by Adam M. Wojciechowski \textit{et al.} \cite{Wojciechowski2018ContributedSensor}, which expresses the sensitivity in terms of the SNR of the image. Under the assumption that optical shot noise dominates, the correlation between the SNR and $\mathcal{N}$ can be written as: 
\begin{equation}
    \sqrt{\mathcal{N}} =SNR\approx\sqrt{S}
\end{equation}
where $S$ is the total signal (in counts) measured during the optical readout. Substituting into equation (\ref{eq:n_p1}), the sensitivity becomes:

\begin{equation}\label{eq:n_p2}
    \eta = \frac{8}{3\sqrt{3}}\frac{\hbar}{g_e\mu_\mathrm{B}}\frac{1}{C\sqrt{S}}\frac{\sqrt{t_\mathrm{o}+T_2^*}}{T_2^*}
\end{equation}

\section{Results}
\begin{figure}[h]
\includegraphics{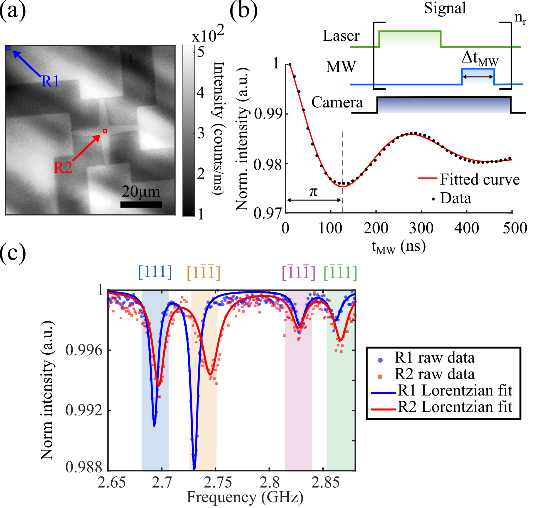}
\caption{\label{fig:3}Experimental Results from NV-Based Magnetometry. (a) Fluorescence image of the two crossing ellipses with two  10 × 10 pixels regions highlighted, blue and red. (b) Protocol implemented to measure the Rabi oscillations (top) and the respective oscillations obtained for the \([1\bar{1}\bar{1}]\) orientation (bottom). The $\pi$ pulse is represented in the figure as the first minimum point of fluorescence corresponding to 110 ns. (c) ODMR spectra for the two regions highlighted in (a). A shift in the resonance frequency is visible between the two areas, influenced by the magnetic field produced by the ellipses.}
\end{figure}

To analyze the vector magnetic field created by the microstructured magnetic sample, we focused on the structure closest to the microwave (MW) antenna and recorded the NV fluorescence for a total exposure time of 20 ms. Figure \ref{fig:3} (a) shows the fluorescence intensity emitted from the NV layer right below two equal crossing ellipses of 40 $\mu m$ x 5 $\mu m$. Light reflections at the diamond–sample interface, combined with the high spatial coherence of our laser, result in interference patterns visible as alternating bright and dark regions.

We first determined the $\pi$-pulse duration using a Rabi experiment (Figure \ref{fig:3} (b)), which yielded a value of $\approx 110$ ns. With this calibration, we implemented the pulsed ODMR protocol across half the spectrum, averaging over 30 repetitions to enhance the signal-to-noise ratio. In Figure \ref{fig:3} (c) we show the ODMR spectrum obtained for two distinct 10 x 10 pixels regions: one inside the magnetic structures (R2) and the other positioned at the top left corner of the FOV (R1). This preliminary analysis reveals that the magnetic field contributions from the ellipses in the overlapping region produce distinct shifts for each NV orientation. The blue ODMR trace, taken from a region far from the permalloy structures, was used to extract the bias field, since no significant contribution from the structures is expected there. Thus the bias field produced by the permanent magnets was determined to be $B_0=(B_x,B_y,B_z)$ = (4.1, 0.72, 1.1) mT. For subsequent magnetic field calculations, this value was subtracted to isolate the contributions from the magnetic structures. After correction, the magnetic field in the R2 region was found to be B=(-0.24, 0.04, 0.28) mT.

\begin{figure}[h]
\includegraphics{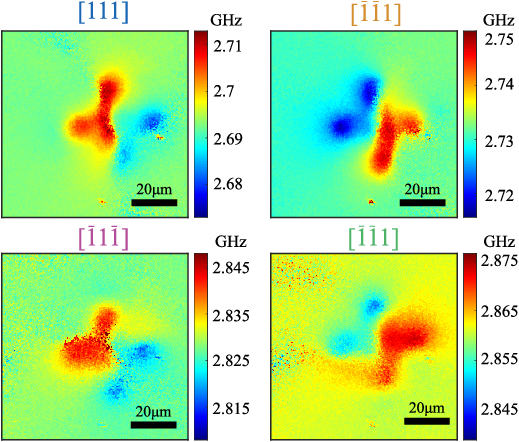}
\caption{\label{fig:3_2}Frequency maps obtained for the four NV orientations ([111],\([1\bar{1}\bar{1}]\), \([\bar{1}1\bar{1}]\), \([\bar{1}\bar{1}1]\)). The frequency variation is confined to the regions inside the ellipses in all cases.}
\end{figure}

To take advantage of widefield imaging, we extracted the resonance-frequency maps by fitting the ODMR spectrum at each pixel after applying a 3 × 3 spatial moving-average to improve SNR. This spatial averaging procedure reduces our optical spatial resolution to  $\approx$ 0.52 $\mu$m since each pixel on the frequency map will correspond to an average of 3 × 3 pixels. Since the ODMR spectrum features four distinct dips corresponding to the NV orientations, the data at each pixel were fitted with a sum of four Lorentzian functions:
\begin{equation}
    L(\nu;A,\nu^{res},\Gamma) = \sum_{i=1}^{4}A_{i}\Big[\frac{\Gamma^2_{i}}{(\nu-\nu^{res}_{i})^2+\Gamma^2_{i}}\Big]
    \label{fit:lorentzian}
\end{equation}
where $A_i$ is the ODMR constrast (dip amplitude), $\Gamma$ is the linewidth, $\nu^{res}_i$ is the resonance frequencies for the i-th NV center orientation, with $i \in [$[111], \([1\bar{1}\bar{1}]\), \([\bar{1}1\bar{1}]\), \([\bar{1}\bar{1}1]\)]. In Fig. \ref{fig:3_2} we demonstrate four frequency maps corresponding to the four different NV orientations ([111], \([1\bar{1}\bar{1}]\), \([\bar{1}1\bar{1}]\), \([\bar{1}\bar{1}1]\)). These maps provide the spatial variation of the resonance frequency value at each pixel for all NV orientations. There is a clear shift in the resonance frequencies in the regions within the magnetic structure, which confirms the capability of our setup to characterize them. However, the resonances with lower contrast ($[\bar{1}1\bar{1}]$, $[\bar{1}\bar{1}1]$) are more difficult to fit accurately due to the reduced signal-to-noise ratio (SNR). This leads to increased variability and reduced magnetic resolution in the corresponding frequency maps. Nevertheless, the projection of the magnetic field was calculated for each orientation at each pixel by subtracting the bias magnetic field from the absolute value.

To quantify the impact of this spatially varying fit quality on the magnetometric performance, we estimated the spatially averaged sensitivity by extracting the ODMR contrast and linewidth from the four corners of the image for each orientation. Averaging across these regions yields mean contrasts of $C = (1.138 \pm 0.042)\%$, $(0.868 \pm 0.041)\%$, $(0.458 \pm 0.008)\%$, and $(0.677 \pm 0.025)\%$, and linewidths of $\Gamma = (4.942 \pm 0.826)$ MHz, $(5.312 \pm 1.667)$~MHz, $(4.961 \pm 1.134)$ MHz, and $(5.169 \pm 0.361)$ MHz, for the orientations $[111]$, $[1\bar{1}\bar{1}]$, $[\bar{1}1\bar{1}]$, and $[\bar{1}\bar{1}1]$, respectively. Using these values in Eq.~\ref{eq:n_p2}, we obtain sensitivities of
$\eta = (828 \pm 142)$, $(1167 \pm 370)$, $(2066 \pm 474)$, and $(1456 \pm 115)$~nT$/\sqrt{\mathrm{Hz}}$.

With the magnetic field projections extracted for all four orientations, the individual magnetic field components can be reconstructed in the laboratory frame. We define our transformation model based on the work of \cite{Wen2024}, which defines the unit vectors according to the laboratory frame (x,y,z) in a diamond with a (100) top surface as:
\begin{equation}\label{eq:lab_frame}
\begin{aligned}
\vec{u}_{[111]}           &= \tfrac{1}{\sqrt{3}}\bigl(\hat{x},\;\hat{y},\;\hat{z}\bigr),\\
\vec{u}_{[1\bar{1}\bar{1}]} &= \tfrac{1}{\sqrt{3}}\bigl(\hat{x},\;-\hat{y},\;-\hat{z}\bigr),\\
\vec{u}_{[\bar{1}1\bar{1}]} &= \tfrac{1}{\sqrt{3}}\bigl(-\hat{x},\;\hat{y},\;-\hat{z}\bigr),\\
\vec{u}_{[\bar{1}\bar{1}1]} &= \tfrac{1}{\sqrt{3}}\bigl(-\hat{x},\;-\hat{y},\;\hat{z}\bigr).
\end{aligned}
\end{equation}

where the edges of the diamond correspond to the (x) and (y) axes, and the (z) axis is defined normal to the
surface. Taking the values of the magnetic field projection along each NV orientation ($B_{[111]}$, $B_{[1\bar{1}\bar{1}]}$, $B_{[\bar{1}1\bar{1}]}$, $B_{[\bar{1}\bar{1}1]}$), the magnetic field components in the lab frame can be calculated as:
\begin{equation}\label{eq:transformation}
\begin{aligned}
B_x &= \tfrac{\sqrt{3}}{4}\bigl(B_{[111]} + B_{[1\bar{1}\bar{1}]} - B_{[\bar{1}1\bar{1}]} - B_{[\bar{1}\bar{1}1]}\bigr),\\
B_y &= \tfrac{\sqrt{3}}{4}\bigl(B_{[111]} - B_{[1\bar{1}\bar{1}]} + B_{[\bar{1}1\bar{1}]} - B_{[\bar{1}\bar{1}1]}\bigr),\\
B_z &= \tfrac{\sqrt{3}}{4}\bigl(B_{[111]} - B_{[1\bar{1}\bar{1}]} - B_{[\bar{1}1\bar{1}]} + B_{[\bar{1}\bar{1}1]}\bigr).
\end{aligned}
\end{equation}

\begin{figure*}[ht]
\includegraphics{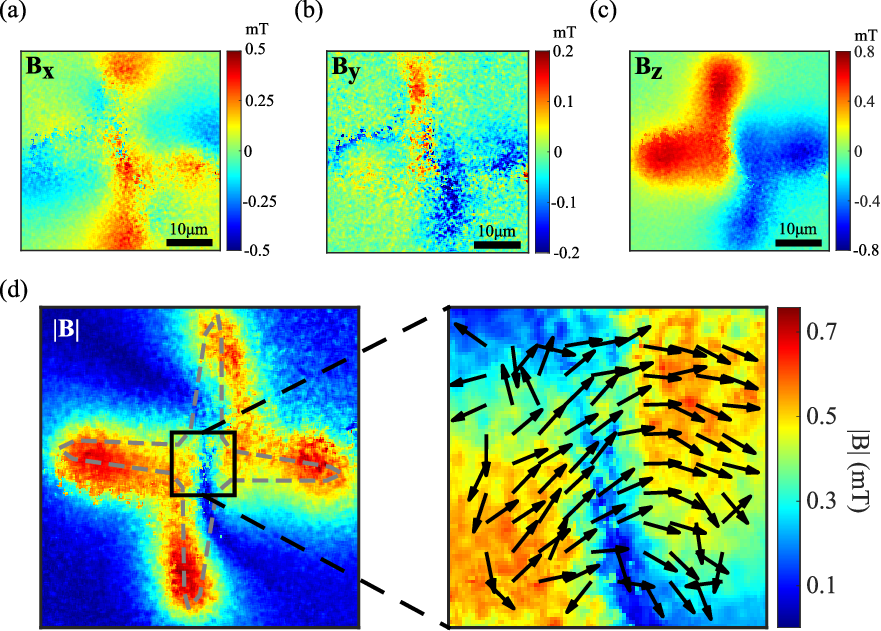}
\caption{\label{fig:4}Maps of the magnetic field vector components: (a) x-component, (b) y-component, and (c) z-component of the magnetic field reconstructed from the four NV orientations. A binning of 3x3 pixels was applied to retain spatial resolution in the reconstructed maps. (d) Map of the magnetic field strength obtained by combining the three vector components. The dashed gray lines delimit the magnetic structure. The inset displays the in-plane field direction in the central region, derived from the $\mathrm{B}_\mathrm{x}$ and $\mathrm{B}_\mathrm{y}$ components.}
\end{figure*}

In Fig. \ref{fig:4} (a)-(c) we show the resulting maps of the individual magnetic field components $B_x$, $B_y$ and $B_z$. For visualization clarity, the images display a 270 × 270-pixel crop of the original 512 × 512 images centered on the structures. This representation highlights the strong out-of-plane magnetic field component ($B_z$) and demonstrates a less well-defined distribution of the field for the $B_x$ and $B_y$ components. To visualize the magnetic field directionality in the xy-plane across the FOV, we combined the components $B_x$ and $B_y$ into a vector field. Figure \ref{fig:4} (d) exhibits the magnitude of the magnetic field, with inset illustrating the field's directionality in the xy-plane at the crossing area of the ellipses. By closely examining the overlapping areas, the magnetic field forms an almost 45º angle at the boundaries between the two domains. This finding confirms the existence of a remanent magnetic state aligned in one of the four expected directions. 

\section{Discussion}
Our work demonstrates the successful implementation of the pulsed ODMR protocol in a widefield imaging configuration to reconstruct the magnetic field from thin structures. To assess the performance, we estimated the minimum detectable magnetic field sensitivity per orientation using equation (\ref{eq:n_p2}). With the  3 × 3 spatial binning we obtained sensitivities in the range of $\eta = $ 0.828 --- 2.07 $\mu T {Hz}^{-1}$. The spread in $\eta$ values reflects an unequal projection of the bias field onto the NV axes, leading to contrast differences between NV center orientations. To the best of our knowledge, this sensitivity is among the highest reported for widefield NV magnetometry in both pulsed and CW ODMR, with previous works not reaching the nanotesla range \cite{Yanling,Sengottuvel,Yoon}. Compared with tip-based or planar scanning probes \cite{Weinbrenner}, our widefield-based method provides a snapshot of the magnetic field distribution across the FOV, extracting vector information while significantly reducing acquisition time and maintaining competitive sensitivity. 

While these results highlight the competitive sensitivity of our approach, there remains room for improvement. Strategies for enhancement include optimization of the optical collection and excitation pathways \cite{Ma,XuL} to exceed the typical 10 \% collection efficiency expected for high NA oil-immersion objectives \cite{LeSage}. Clevenson \textit{et al.} \cite{Clevenson} improved the pump absorption efficiency by implementing light-trapping techniques that resulted in an increase of more than two orders of magnitude over previous single-pass schemes, and Yu \textit{et al.} \cite{Yu} achieved a 92\% increase in collection efficiency by combining reflective coatings with a compound parabolic concentrator. Further sensitivity enhancements are possible through improvements of the antenna design and fabrication, which can deliver more homogeneous, higher-amplitude MW fields and greater control of spin states over larger areas \cite{Opaluch,LiZ}. Replacing the permanent magnets with Helmholtz coils for bias field generation would also provide magnetic fields that are more homogeneous and controllable. 

Additionally, we observe a blur of the magnetic field around the ellipse's boundaries and a reduction in the sharpness of the magnetic field distribution. This is mainly caused by an undesirable standoff distance between the magnetic sample and the diamond sensing platform that leads to spatial averaging effects \cite{Levine}. This issue can be mitigated by optimizing the interface between the sample and the diamond, for example, by gluing them together or by using a specific holder that enables precise and repeatable sample-sensor interfacing, as demonstrated by K.J. Rietwyk \textit{et al.} \cite{Rietwyk2024PracticalMicroscope}.

Beyond our achieved sensitivity, the main advantage of our approach is the speed for magnetic field characterization achieved by recording a large area simultaneously, providing ODMR traces in every image pixel of the FOV. In our implementation, this parallelism yields a total acquisition time of 4.62 mins for a full vector map, whereas single-NV scanning or confocal approaches would require substantially longer times for comparable scanning areas. The imaging speed could be further increased by multiplexing the four NV orientations in a single acquisition with or without lock-in detection, as demonstrated by Schloss \textit{et al.} \cite{Schloss2} and Shi Z. \textit{et al.} \cite{Shi}. In addition, Parashar \textit{et al.} \cite{Parashar} demonstrated that widefield per-pixel lock-in detection of frequency-modulated NV-ODMR enables high-frame-rate imaging, allowing for the study of dynamically changing fields.

To validate the assignment of NV axes to the measured resonance frequencies, the experiments could be repeated after preparing the magnetic field in a known state through planar Hall response measurements. Alternatively, simulations of the magnetic field projection at different distances from the magnetic material surface (z value) could provide a theoretical confirmation of our experimental findings. 

Potential applications include the visualization of two-dimensional spin textures such as magnetic skyrmion bubbles \cite{Dohi} and the study of van der Waals magnets \cite{YangS,Robertson} as well as current imaging of integrated-circuitry \cite{Garsi}. Moreover, with the improvements in sensitivity and speed, our approach could be used to track the magnetization switching dynamics in artificial synapses \cite{ZhangF}, in thin-films \cite{Telepinsky2016}, and other emerging spintronic platforms that currently lack suitable magnetic imaging techniques for detailed characterization.
\section{Conclusion\label{except}}
In conclusion, we successfully mapped stray vector magnetic fields produced by magnetic thin-film structures using a pulsed ODMR protocol with nanotesla sensitivity and fast acquisition times. Using 3 × 3 spatial binning, we achieved per-orientation sensitivities of $\eta = $ 0.828 --- 2.07 $\mu T Hz^{-1}$, largely homogeneous across the field of view, and obtained vector maps within minutes. By combining the speed of widefield imaging with the sensitivity of pulsed protocols, our method overcomes many of the limitations associated with scanning probe techniques, offering a scalable approach to spatially resolved vector magnetometry. Future improvements in optical collection, antenna design, and sensor-sample interfacing could push the sensitivity towards the picotesla range and broaden applicability across diverse systems. In future work, we plan to employ this platform for the imaging of magnetic vortex structures and to study the variability of the vortex center position under different external magnetic fields. Overall, the compatibility with standard optical microscopy makes this approach both scalable and accessible for investigating magnetic phenomena in both biological and material science.
\begin{acknowledgments}
We thank Beatriz Costa for the RF antennas that we could use, which were characterized at the spintronics laboratories of INL. We counted on the access and support of the INL Nanophotonics and Bioimaging (NBI) Research Core Facilities and thank Mariana Carvalho for support in the adaptation of a commercial widefield system for ODMR upgrade. This work received funding by the FCT (Fundação para a Ciência e Tecnologia) via the project DIAMOND-CONNECT with grant agreement ID PTDC/NAN-OPT/7989/2020 and by La Caixa Foundation (ID 100010434) and FCT via the project Diamond4Brain with grant agreement ID LCF/PR/HP20/52300001. J.P.S. acknowledges an individual PhD fellowship from the FCT with grant no.: 2022.11803.BD. A Seed grant provided by host institutions INL and BINA for the project entitled “Directional quantum sensing based on nitrogen vacancy centers in diamond for magnetic field mapping of multi-level magnetic memory structures" funded by the joint call in 2021-2023.
\end{acknowledgments}

\section*{Data availability}
The data that support the findings of this study are available upon reasonable request from the authors.

\bibliography{references}

\end{document}